\documentstyle[12pt]{article}

\begin{document}

\title{ Summation formula for solutions of Riccati-Abel equation }
\author{
Robert M. Yamaleev\\
Joint Institute for Nuclear Research, LIT, Dubna, Russia.\\
Universidad Nacional Autonoma de Mexico, Mexico.\\
Email: yamaleev@jinr.ru } \maketitle
%
% Abstract
%
\begin{abstract}

The Riccati-Abel equation defined as an equation between first
order derivative and cubic polynomial is explored. In the case of
constant coefficients this equation is reduced into algebraic
equation. The method of derivation of a summation formula for
solutions of Riccati-Abel equation elaborated. Interrelation with
general complex algebra of third order is established.

\end{abstract}

\section{ Introduction}

Consider the first order differential equation
$$
f(u,x)=\frac{du}{dx}.  \eqno(1.1)
$$
If we approximate $f(u,x)$, while $x$ is kept constant, we will
get
$$
Q_0(x)+Q_1(x)u+Q_2(x)u^2+Q_3(x)u^3+\cdots=\frac{du}{dx}.
\eqno(1.2)
$$
When the series in the left-hand side is restricted with second
order polynomial the equation is the Riccati equation
\cite{Riccati}.

The Riccati equation is one of the widely used equations of
mathematical physics.  The ordinary Riccati equations are closely
related with second order linear differential equations. For the
solutions of the ordinary Riccati equations with constant
coefficients a summation formula can be derived. These solutions
are presented by trigonometric functions induced by general
complex algebra.

 If, in particular, $f(u,x)$ is a cubic polynomial,
then the equation is called Riccati-Abel equation. Abel's original
equation was written in the form
$$
(y+s)\frac{dy}{dx}+p+qy+ry^2=0. \eqno(1.3)
$$
This equation is converted into Riccati-Abel equation by
transformation $y+s=1/z$, which yields
$$
\frac{dz}{dx}=rz+(q-s'-2rs)z^2+(p-qs+rs^2)z^3.  \eqno(1.4)
$$
It is seen that the case $Q_0(u,\phi)=0$ was  actually considered
by Abel \cite{Abel}.

When the series in the left-hand side of equation (1.2) is given
by $n$-order polynomial we deal with the generalized Riccati
equations. The solution of the generalized Riccati equation with
constant coefficients can be denominated as {\it generalized
tangent function}. The generalized Riccati equations are used, for
example, in various problems of renorm-group theory
\cite{Yeomans}. The mean field free energy concept and the
perturbation renormalization group theory deal with differential
equations of first order with polynomial non-linearity.

 The aim of this paper is to
 explore solutions of the Riccati-Abel equation with constant
 coefficients and to derive some kind of summation formula
for them. Summation (addition) formulae for solutions of linear
differential equations are considered as important features of
these functions. Let us mention, for example, summation formulae
for the trigonometric sine-cosine functions, the Bessel functions,
the hypergeometric functions and their various generalizations.
Whereas solutions of the linear differential equations with
constant coefficients  admit universal methods of obtaining of
summation formulas (see, for instance, \cite{Ungar1},
\cite{Ungar2}), the solutions of nonlinear equations require
special investigations. In this context let us mention the
addition formulae for Jacobi and Weierstrass elliptic functions
\cite{Akhiezer}.

 The solutions of the generalized Riccati
equations with cubic and higher polynomials,
 in general, do not admit any summation formula.
Nevertheless, by careful analysis we found a new summation law
according to which in order to obtain a summation formula for the
solutions of the third order Riccati equation is necessary to use
two independent variables. We will show that the summation formula
can be derived also by using interconnection between solutions of
Riccati-Abel equation and the characteristic functions of
generalized complex algebra of third order.

The paper is presented by the following sections.  Section 2
committed to solution of ordinary Riccati equation with constant
coefficients. Summation formula for the solutions are derived and
interrelation with solutions of the linear differential equations
is underlined. In Section 3, the Riccati-Abel equation is
integrated, the corresponding algebraic equation for solutions is
derived, a summation formula for solutions is established. In
Section 4, the solutions of Riccati-Abel equation are constructed
within generalized complex algebra of third order.

\section{ Ordinary Riccati equation, summation formula and general complex algebra }

{\bf 2.1 The ordinary Riccati equation}.

Consider the Riccati equation with constant coefficients
$$
u^2-a_1u+a_0=\frac{du}{d\phi}.       \eqno(2.1)
$$
If coefficients $a_0,a_1$ are constants then a great
simplification results because it is possible to obtain the
complete solution by means of quadratures.
 Thus, equation (2.1) admits direct integration
$$
\int~\frac{dx}{x^2-a_1x+a_0} =\int d\phi.       \eqno(2.2)
$$
 Let $x_1,x_2 \in C$ be roots of the polynomial equation
$$
x^2-a_1x+a_0=0.           \eqno(2.3)
$$
In order to calculate the integral (2.2) the following formula
expansion is used
$$
\frac{1}{x^2-a_1x+a_0}=\frac{1}{2x_1-a_1}\frac{1}{x-x_1}+
\frac{1}{2x_2-a_1}\frac{1}{x-x_2},   \eqno(2.4)
$$
where,
$$
{2x_1-a_1}=(x_1-x_2),~~{2x_2-a_1}=(x_2-x_1).
$$
Then the integral (2.2) is easily calculated and the result is
given by the logarithmic functions
$$
\int^u_w~\frac{dx}{x^2-a_1x+a_0}=
\frac{1}{m_{12}}(~\log\frac{u-x_1}{u-x_2}
-~\log\frac{w-x_1}{w-x_2}~)=\phi(u)-\phi(w),            \eqno(2.5)
$$
where $ m_{12}=x_1-x_2.$
 Now, let us keep the first logarithm of (2.5) depending of the initial limit of the integral, that is
$$
\frac{1}{m_{12}}\log[\frac{u-x_1}{u-x_2}] =\phi(u).
$$
By inverting the logarithm function we come to the algebraic
equation for solution of (2.1),
$$
\exp(m_{12}\phi)=\frac{u-x_1}{u-x_2}.  \eqno(2.6)
$$
 Let $ u(\phi_0)=0$, then
$$
\exp(m_{12}\phi_0)=\frac{x_1}{x_2}.\eqno(2.7)
$$
As soon as the point $\phi=\phi_0 $ is determined, one may
calculate the function $u(\phi)$ by making use of algebraic
equation (2.6). Since $a_1=x_1+x_2$, from (2.7) it follows that
$$
a_1=m_{12}\coth(m_{12}\phi_0/2).
$$
Consequently, from (2.6) we obtain
$$
u(\phi,\phi_0)=\frac{1}{2}m_{12}\coth(m_{12}\phi_0/2)-\frac{1}{2}m_{12}\coth(m_{12}\phi/2).
$$

{\bf 2.2 Summation (addition) formula for function
$u=u(\phi,\phi_0)$.}

Consider the following integral equation
$$
\int^u~\frac{dx}{x^2-a_1x+a_0}+\int^v~\frac{dx}{x^2-a_1x+a_0}=\int^w~\frac{dx}{x^2-a_1x+a_0}.
    \eqno(2.8)
$$
The quantity $w$ is a function of $u$ and $v$, if the function
$w=f(u,v)$ is an algebraic function then this function can be
considered as the summation formula. Write (2.8) in the following
notations $\phi_u+\phi_v=\phi_w$. Then,
$w(\phi_w)=w(\phi_u+\phi_v)=f(u(\phi_u),~v(\phi_v)~).$

Calculating the integrals in (2.8) we come to the following
algebraic equation
$$
\frac{1}{2m}\log\frac{u-x_1}{u-x_2}\frac{v-x_1}{v-x_2}=\frac{1}{2m}\log\frac{w-x_1}{w-x_2}.
\eqno(2.9)
$$
Thus, the function $w(u,v)$ has to satisfy the equation
$$
\frac{u-x_1}{u-x_2}\frac{v-x_1}{v-x_2}=\frac{w-x_1}{w-x_2}.
\eqno(2.10)
$$
Multiplying fractions and taking into account the fact that
$x_1,x_2$ obey (2.3), we get
$$
\frac{uv-x_1(u+v)+a_1x_1-a_0}{uv-x_2(u+v)+a_1x_2-a_0}=
$$
$$
=\frac{\frac{uv-a_0}{u+v-a_1}- x_1}{\frac{uv-a_0}{u+v-a_1}-
x_2}=\frac{w-x_1}{w-x_2},
$$
$$
w=\frac{uv-a_0}{u+v-a_1}.\eqno(2.11)
$$
This is the summation formula for function $u(\phi;a_1,a_0)$.

{\bf 2.3 Relationship with General complex algebra}.

Like the (co)tangent function can be defined  as a ratio of cosine
and sine functions, the solution of the Riccati equation
$u(\phi;a_1,a_0)$ also can be represented as a ratio of modified
cosine and sine functions. Firstly, let us construct these
functions.

Consider  general complex algebra generated by the $(2\times 2)$
matrix \cite{Yamaleev1}
$$
E=\left( \begin{array}{cc}
0&-a_0\\
1&a_1/2
\end{array} \right) \eqno(2.12)
$$
obeying the quadratic equation (2.3):
$$
E^2-a_1E+a_0I=0,  \eqno(2.13)
$$
with $I$- unit matrix. Expansion with respect to $E$ of the
exponential function $\exp(E\phi)$ leads to the Euler formula
\cite{Babusci}
$$
\exp(E\phi)=g_1(\phi;a_0,a_1)E+g_0(\phi;a_0,a_1).  \eqno(2.14)
$$
In terms of the roots $x_1,x_2$ this matrix equation is separated
into two equations
$$
\exp(x_2\phi)=x_2~g_1(\phi;a_0,a_1)+g_0(\phi;a_0,a_1),~~
\exp(x_1\phi)=x_1~g_1(\phi;a_0,a_1)+g_0(\phi;a_0,a_1), \eqno(2.15)
$$
from which an explicit form of $g$-functions can be obtained.
Apparently, $g_0$ and $g_1$ are modified (generalized) cosine-sine
functions with the following formulas of differentiation
$$
\frac{d}{d\phi} g_1(\phi;a_0,a_1)=
g_0(\phi;a_0,a_1)+a_1~g_1(\phi;a_0,a_1),~~ \frac{d}{d\phi}
g_0(\phi;a_0,a_1)=-a_0~g_1(\phi;a_0,a_1). \eqno(2.16)
$$
Form a ratio of two equations of (2.15) as follows
$$
\exp(m_{21}\phi)=\frac{x_2~g_1(\phi;a_0,a_1)+g_0(\phi;a_0,a_1)}{x_1~g_1(\phi;a_0,a_1)+g_0(\phi;a_0,a_1)}.
\eqno(2.17)
$$
Let $g_1(s;a_0,a_1)\neq 0$. Then,
$$
\exp(m_{21}\phi)=\frac{x_2+D}{x_1+D},  \eqno(2.18)
$$
where
$$
D=\frac{g_0(\phi;a_0,a_1)}{g_1(\phi;a_0,a_1)}.
$$
Differential equation for function $D(\phi)$ is obtained by using
(2.16):
$$
D^2+a_1D+a_0=-\frac{dD}{d\phi}. \eqno(2.19)
$$
 Thus, we have proved that the function
$$
u(\phi;a_0,a_1)=-D=-\frac{g_0(\phi;a_0,a_1)}{g_1(\phi;a_0,a_1)}
\eqno(2.20)
$$
obeys the Riccati equation.

The summation formulae for $g$-functions are well defined
\cite{Yamaleev1}. They are
$$
g_0(a+b)=g_0(a)g_0(b)-a_0g_1(a)g_1(b),~~
$$
$$
g_1(a+b)=g_1(a)g_0(b)+g_0(a)g_1(b)+a_1g_1(a)g_1(b).
$$
$$
\frac{g_0(a+b)}{g_1(a+b)}=\frac{g_0(a)g_0(b)-a_0g_1(a)g_1(b)}{g_1(a)g_0(b)+g_0(a)g_1(b)+a_1g_1(a)g_1(b)}.
\eqno(2.21)
$$
By taking into account (2.20) we get
$$
u(a+b)= -\frac{g_0(a+b)}{g_1(a+b)}=
\frac{u(a)u(b)-a_0}{u(a)+u(b)-a_1}.  \eqno(2.22)
$$
which coincides with (2.11).

\section{ Generalized Riccati  equation with cubic order polynomial }

{\bf 3.1 The Riccati-Abel equation}.

Consider the following non-linear differential equation with
constant coefficients
$$
u^3-a_2u^2+a_1u-a_0=\frac{du}{d\phi},       \eqno(3.1)
$$
which admits direct integration by
$$
\int^u_w~\frac{dx}{x^3-a_2x^2+a_1x-a_0}=\phi(w)-\phi(u).
\eqno(3.2)
$$
This integral is calculated by making use of well-known method of
partial fractional decomposition \cite{fraction}
$$
\frac{1}{x^3-a_2x^2+a_1x-a_0}=\frac{1}{(x-x_3)(x-x_2)(x-x_1)}=
$$
$$
\frac{(x_3-x_2)}{V}\frac{1}{x-x_1}+\frac{(x_1-x_3)}{V}\frac{1}{x-x_2}+\frac{(x_2-x_1)}{V}\frac{1}{x-x_3},
\eqno(3.3)
$$
where $V$ is the Vandermonde's determinant \cite{Vein}
$$
V=(x_1-x_2)(x_2-x_3)(x_3-x_1),\eqno(3.4)
$$
and the distinct constants $x_1,x_2,x_3 \in C$ are roots of the
cubic polynomial
$$
f(x)=x^3-a_2x^2+a_1x-a_0=0. \eqno(3.5)
$$
By using expansion (3.3) the integral (3.2) is easily calculated
$$
\int^u_w~\frac{dx}{x^3-a_2x^2+a_1x-a_0}=
$$
$$
\frac{(x_3-x_2)}{V}\log\frac{u-x_1}{w-x_1}+\frac{(x_1-x_3)}{V}\log\frac{u-x_2}{w-x_2}+
\frac{(x_2-x_1)}{V}\log\frac{u-x_3}{w-x_3}=\phi(u)-\phi(w).
\eqno(3.6)
$$
Introduce the following notations
$$
m_{ij}=(x_i-x_j),~i,j=1,2,3,~\mbox{with}~m_{21}+m_{32}+m_{13}=0.
\eqno(3.7)
$$
Equation (3.6) re-write as follows
$$
\int^u~\frac{dx}{x^3-a_2x^2+a_1x-a_0}=
\log{~(u-x_1)^{m_{32}}(u-x_2)^{m_{13}}(u-x_3)^{m_{21}}}=V\phi(u)
\eqno(3.8)
$$
and invert the logarithm, this leads to the following algebraic
equation
$$
[u-x_1]^{m_{32}}[u-x_2]^{m_{13}}[u-x_3]^{m_{21}}=\exp(V\phi).
\eqno(3.9)
$$
This equation can be written also in the fractional form
$$
[\frac{u-x_1}{u-x_3}]^{m_{32}}[\frac{u-x_2}{u-x_3}]^{m_{13}}~=\exp(V\phi).\eqno(3.10)
$$
 Thus, the problem of solution of differential (3.1) is reduced to the problem of solution of the algebraic equation (3.10).
 Notice, if the roots of cubic equation and function $u$ are defined in the field of real numbers then this equation is meaningful only for
 certain domain of definition of $u(\phi)$.

{\bf 3.2 Semigroup property of fractions of $n$-order monic
polynomials on the set of roots of $n+1$-order polynomial.}

In this section let us recall semigroup property of the fractions
of $n$-order polynomials defined on the set of roots of
$n+1$-order polynomial.
 Let $F(x,n+1)$ be $(n+1)$ order polynomial
with $(n+1)$ distinct roots $x_i,i=1,\ldots,n+1$. Denote this set
of roots by $FX(n+1)$.

{\bf Lemma 3.1}

Let $P_a(x_i,n)$ be $n$-order polynomial on $x_i\in FX(n+1)$. The
product of two $n$-order polynomials
$$
P_a(x_i,n)*P_b(x_i,n)
$$
is also $n$-order polynomial $P_c(x_i,n)$.

{\bf Proof}.

The product $P_{ab}(x_i,2n):=P_a(x_i,n)*P_b(x_i,n)$ is polynomial
of $2n$-degree with respect to variable $x_i$. Since $x_i$ obeys
$n+1$-order polynomial equation all degrees of the variable $x_i$
from $(n+1)$ up till $2n$ degree can be expressed via $n$-degree
polynomial. In this way the polynomial $P_{ab}(x_i,2n)$ is reduced
till $n$ degree polynomial with respect to variable $x_i\in
FX(n+1)$.

{\bf End of proof}.

are $n$-order monic polynomials defined on the set of the roots of
polynomial $F(x,n+1)$.

Consider two monic polynomials of $n$-degree
$P_a(x_i,n),~P_b(x_k,n)$ with $x_i\neq x_k\in FX(n+1)$. Form a
rational algebraic fraction
$$
\frac{P_a(x_i,n)}{P_a(x_k,n)}.
$$
 The following {\bf Corollary 3.2} holds true:

{\it The product of two fractions formed by two $n$-order monic
polynomials on the roots of $(n+1)$-order polynomial is a fraction
of the same order monic polynomials on the variables},
$$
\frac{P_a(x_i,n)}{P_a(x_k,n)}\frac{P_b(x_i,n)}{P_b(x_k,n)}=\frac{P_c(x_i,n)}{P_c(x_k,n)}.
$$

{\bf 3.3 Addition formula for  $u(\phi)$. }

 Let $\phi=\phi_0$ be a point where $u(\phi_0)=0$. Then, (3.10)
  is reduced to
$$
[\frac{x_1}{x_3}]^{m_{32}}[\frac{x_2}{x_3}]^{m_{13}}=\exp(V\phi_0).
\eqno(3.11)
$$
If we make simultaneous translations of the roots $x_k,k=1,2,3$ by
some value $u$ in the left-hand side of (3.11), then in the
right-hand side of the equation $V$ does not change, hence
$\phi_0$ will undergo some translation by $\phi=\phi_0+\delta$. In
this way one may construct the solution of Riccati-Abel equation
(3.1) with initial condition $u(\phi_0)=0.$

Now, let $u,v,w$ be solutions of equation (3.1) calculated for
tree variables $\phi_u,\phi_v,\phi_w$, which obey the equation
$\phi_w=\phi_u+\phi_v$. Then, in accordance with (3.10) we write
$$
\exp(V\phi_u)\exp(V\phi_v)
=\{~[\frac{u-x_1}{u-x_3}\frac{v-x_1}{v-x_3}]^{m_{32}}
[\frac{u-x_2}{u-x_3}\frac{v-x_2}{v-x_3}]^{m_{21}}~~\}
$$
$$
=\{~(\frac{w-x_1}{w-x_3})^{m_{32}}[\frac{w-x_2}{w-x_3}]^{m_{21}}~~\}
=\exp(V(\phi_u+\phi_v).     \eqno(3.12)
$$
The problem is to find some rational function expressing $w$ via
the pair $(u,v)$, i.e., the function $w=w(u,v)$ has to be a
rational function.

 Evidently,  the method used in the previous section for the
 ordinary Riccati equation now is not applicable. According to Lemma
we are able to transform a product of ratios of $n$-order
polynomials into the ratio of $n$-order polynomials if these
polynomials are defined on roots of $n+1$-order polynomial. Thus,
we have to seek another way of construction of a summation
formula.

{\it The problem we suggest to resolve as follows.}

Let us to present the integral (3.8) as a sum of two integrals by
$$
\int^w~\frac{dx}{x^3-a_2x^2+a_1x-a_0}=
\int^u~\frac{dx}{x^3-a_2x^2+a_1x-a_0}+
\int^v~\frac{dx}{x^3-a_2x^2+a_1x-a_0} =
$$
$$
=\phi=V\log(~(~\frac{u-x_1}{u-x_3}~\frac{v-x_1}{v-x_3}~)^{m_{32}}
~(~\frac{u-x_2}{u-x_3}~\frac{v-x_2}{v-x_3}~)^{m_{13}}~).
\eqno(3.13)
$$
In this way we arrive to the following algebraic equation
$$
~(~\frac{u-x_1}{u-x_3}~\frac{v-x_1}{v-x_3}~)^{m_{32}}
(~\frac{u-x_2}{u-x_3}~\frac{v-x_2}{v-x_3}~)^{m_{13}}~~)=\exp(V\phi(u,v))=\exp(V\phi_u)\exp(V\phi_v).
\eqno(3.14)
$$
Let $u,v$ be solutions of the quadratic equation
$$
x^2+tx+s=0,~t=-(u+v),~s=uv. \eqno(3.15)
$$
Then, equation (3.14) is written as
$$
[~\frac{x_1^2+tx_1+s}{x_3^2+tx_3+s}~]^{m_{32}}
*(~\frac{x_2^2+tx_2+s}{x_3^2+tx_3+s}~~)^{m_{13}}~~)=\exp(V\phi(t,s)).
\eqno(3.16)
$$
Thus from the pair of functions $(u,v)$ we come to another pair
$(t,s)$.  This pair of functions, in fact, admits a summation rule
because the problem is reduced to the task of transformation
four-degree polynomial into quadratic polynomial at the solutions
of the cubic equation. Evidently, this task can be easily
performed by simple algebraic operations.\\

{\bf Theorem 3.3}.

{\it The following  summation formula for solutions of
Riccati-Abel equation holds true}
$$
(t,s)\bigoplus (v,u)=(r,w),
$$
where
$$
r=~\frac{~(a_0-2a_2a_1)-a_1(v+t)+(tu+sv)}{(~3a_2^2-a_1)+a_2(v+t)+(s+u+tv)}~
~w=\frac{a_2a_0+(v+t)a_0+su}{(~3a_2^2-a_1)+a_2(v+t)+(s+u+tv)}.
\eqno(3.17)
$$

{\bf Proof.}

 Consider product of two monic polynomials
$$
(x^2+tx+s)(x^2+vx+u)=x^4+x^3(v+t)+x^2(s+u+tv)+x(tu+vs)+su,
$$
where $x$ is one of roots of cubic equation
$$
x^3-a_2x^2+a_1x-a_0=0. \eqno(3.18)
$$
From the cubic equation (3.18) we are able to express $x^3$ and
$x^4$ as polynomials of second order as follows
$$
x^3=a_2x^2-a_1~x+a_0,~~x^4=(~3a_2^2-a_1)x^2+(a_0-a_1a_2)~x+a_2a_0.
$$
Then, four-degree polynomial on roots of the cubic polynomial is reduced into polynomial of second order
$$
x^4+x^3(v+t)+x^2(s+u+tv)+x(tu+vs)+su= Ax^2+Bx+C, \eqno(3.19)
$$
where
$$
A=(~3a_2^2-a_1)+a_2(v+t)+(s+u+tv),~B=(a_0-2a_2a_1)-a_1(v+t)+(tu+sv),~
C=a_2a_0+(v+t)a_0+su. \eqno(3.20)
$$
Since we deal with the ratios of polynomials the coefficients of
the quadratic polynomial in (3.19) and polynomials in denominator
and in numerator have the same leading coefficient, we are able
return to the ratio of monic polynomials.
 In this way we come to the relations
$$
r=\frac{B}{A},~~w=\frac{C}{A}. \eqno(3.21)
$$

{\bf End of proof}.

\section{  Generalized complex algebra of third order and solutions of Riccati-Abel equation}

In this section we will establish a relationship between
characteristic functions of general complex algebra of third order
and solutions Riccati-Abel equation.

The unique generator $E$ of general complex algebra of
third-order, $CG_3$, is defined by cubic equation \cite{Yamaleev2}
$$
E^3-a_2E^2+a_1E-a_0=0.\eqno(4.1)
$$
The companion matrix $E$ of the cubic equation (4.1) is given by
$(3\times 3)$ matrix
$$
E:=\left( \begin{array}{ccc}
0&0&a_0\\
1&0&-a_1\\
0&1&a_2
\end{array} \right).\eqno(4.2)
$$
Consider the expansion
$$
\exp(E\phi_1+E^2\phi_2)=g_0(\phi_1,\phi_2)+E~g_1(\phi_1,\phi_2)+E^2~g_2(\phi_1,\phi_2).
\eqno(4.3)
$$
This is an analogue of the Euler formula for exponential function,
the function $g_0(\phi_1,\phi_2)$ is an analogue of cosine
function, and $g_k(\phi_1,\phi_2),k=0,1,2$ are extensions of the
sine function. It is seen, the characteristic functions of $GC_3$
algebra depend of pair of "angles". Correspondingly, for each of
them we have {\it formulae of differentiation}.
$$
\frac{\partial}{\partial\phi_1} \left( \begin{array}{c}
g_0\\
g_1\\
g_2
\end{array}\right)=
\left( \begin{array}{ccc}
0&0&a_0\\
1&0&-a_1\\
0&1&a_1
\end{array} \right)
\left( \begin{array}{c}
g_0\\
g_1\\
g_2
\end{array} \right),
\eqno(4.4)
$$

$$
\frac{\partial}{\partial\phi_2} \left( \begin{array}{c}
g_0\\
g_1\\
g_2
\end{array} \right)=
\left( \begin{array}{ccc}
0&a_0&a_0a_2\\
0&-a_1& a_0-a_1a_2\\
1&a_2&-a_1+a_2^{2}
\end{array} \right)
\left( \begin{array}{c}
g_0\\
g_1\\
g_2
\end{array} \right).
\eqno(4.5)
$$

The semigroup of multiplications of the exponential functions
leads to the following {\it the  addition formulae for $g$-
functions}\cite{Yamaleev3}
$$
\left( \begin{array}{c}
g_0\\
g_1\\
g_2
\end{array} \right)_{(\psi_c=\psi_a+\psi_b)}=
\left( \begin{array}{ccc}
g_0&g_2a_0&g_1a_0+g_2a_0a_2\\
g_1&g_0-g_2a_1&-g_1a_1+g_2( a_0-a_1a_2)\\
g_2&g_1+g_2a_2&g_0+g_1a_2+g_2(-a_1+a_2^{2})
\end{array} \right)_{\psi_a}
\left( \begin{array}{c}
g_0\\
g_1\\
g_2
\end{array} \right)_{\psi_b},
\eqno(4.6)
$$
where the sub-indices of the brackets indicate  dependence of the
$g$-functions of the pair of variables
$\psi_i=(\phi_{1i},\phi_{2i}),i=a,b,c$.

Introduce two fractions of $g$-functions by
$$
tg=\frac{g_1}{g_2},~~sg=\frac{g_0}{g_2}.  \eqno(4.7)
$$
It is seen, these functions are analogues of tangent-cotangent
functions. From addition formulae for $g$-functions (4.6) the
following summation formulae for the general tangent functions are
derived.
$$
t_0=g_0/g_2,~~r_0=f_0/r_2. \eqno(4.8)
$$
$$
T_0=\frac{t_0r_0+a_0(r_1+t_1)+a_0a_2}{
r_0+(t_1+a_2)r_1+t_0+t_1a_2+(-a_1+a_2^{2})~},\eqno(4.9)
$$
$$
T_1=\frac{t_1r_0+t_0r_1-a_1(r_1+t_1)+(
a_0-a_1a_2)~)}{r_0+t_0+a_2(t_1+r_1)+t_1r_1+(-a_1+a_2^{2})~}.
\eqno(4.10)
$$
Here the following notations are used
$$
T_0(\psi_c)=\frac{g_0(\psi_c)}{g_2(\psi_c)},~~T_1(\psi_c)=\frac{g_1(\psi_c)}{g_2(\psi_c)},
$$
$$
t_0(\psi_a)=\frac{g_0(\psi_a)}{g_2(\psi_a)},~
r_0(\psi_b)=\frac{g_0(\psi_b)}{g_2(\psi_b)},
$$
$$
t_1(\psi_a)=\frac{g_1(\psi_a)}{g_2(\psi_a)},~
r_1(\psi_b)=\frac{g_1(\psi_b)}{g_2(\psi_b)},
$$
and $\psi_i=(\phi_{1i},\phi_{2i}),i=a,b$,
$\psi_c=(\phi_{1c}=\phi_{1a}+\phi_{1b},\phi_{2c}=\phi_{2a}+\phi_{2b})$.

Let $x_1,x_2,x_3 \in C$ be eigenvalues of $E$ given by distinct
values. Then, the matrix equation (4.3) is represented by three
separated series $(k=1,2,3)$:
$$
\exp(x_k\phi_1+x_k^2\phi_2)=g_0(\phi_1,\phi_2)+x_k~g_1(\phi_1,\phi_2)+x_k^2~g_2(\phi_1,\phi_2),\eqno(4.11)
$$
Form the following ratios $i\neq k$:
$$
\exp((x_i-x_k)\phi_1+(x_i^2-x_k^2)\phi_2)=\frac{g_0(\phi_1,\phi_2)+x_i~g_1(\phi_1,\phi_2)+x_i^2~g_2(\phi_1,\phi_2)}{
g_0(\phi_1,\phi_2)+x_k~g_1(\phi_1,\phi_2)+x_k^2~g_2(\phi_1,\phi_2)~}.\eqno(4.12)
$$
Consider two of these ratios, namely,
$$
\exp(m_{13}\phi_{1}+(x_1^2-x_3^2)\phi_2)=\frac{g_2x^2_1+g_1x_1+g_0}
{g_2x^2_3+g_1x_3+g_0},  \eqno(4.13a)
$$
$$
\exp(m_{23}\phi_{1}+(x_2^2-x_3^2)\phi_2)=\frac{g_2x^2_2+g_1x_2+g_0}
{g_2x^2_3+g_1x_3+g_0}.  \eqno(4.13b)
$$
where $m_{ij}=x_i-x_j$. The both sides of equation (4.13a) raise
to power $m_{32}$ and the both sides of equation (4.13b) raise to
power $m_{13}$ and multiply left and right sides of the obtained
equations, correspondingly. And, by taking into account that
$m_{13}m_{32}+m_{23}m_{13}=0$, we arrive to the following equation
$$
\exp(m_{13}m_{32}\phi_{1}+(x_1+x_3)m_{13}m_{32}\phi_2)~
\exp(m_{23}m_{13}\phi_{1}+(x_2+x_3)m_{13}\phi_2)
$$
$$
=[\frac{g_2x^2_2+g_1x_2+g_0} {g_2x^2_3+g_1x_3+g_0}]^{m_{13}} ~
[\frac{g_2x^2_1+g_1x_1+g_0} {g_2x^2_3+g_1x_3+g_0}]^{m_{32}}.
\eqno(4.14)
$$
The left hand side of this equation is equal to $\exp(V\phi_2)$,
that is,
$$
\exp(V\phi_2)=[\frac{g_2x^2_2+g_1x_2+g_0}{g_2x^2_3+g_1x_3+g_0}]^{m_{13}}
~ [\frac{g_2x^2_1+g_1x_1+g_0} {g_2x^2_3+g_1x_3+g_0}]^{m_{32}}.
\eqno(4.15)
$$
Let $g_2\neq 0$, then by dividing numerator and denominator by
$g_2$ we obtain
$$
\exp(V\phi_2)=[\frac{x^2_2+tg~x_2+sg}{x^2_3+tg~x_3+sg}]^{m_{13}} ~
[\frac{x^2_1+tg~x_1+sg} {x^2_3+tg~x_3+sg}]^{m_{32}}, \eqno(4.16)
$$
where
$$
tg=\frac{g_1}{g_2},~~sg=\frac{g_0}{g_2}.
$$

Let $u,v$ be roots of the quadratic equation
$$
g_0(\phi_1,\phi_2)+y~g_1(\phi_1,\phi_2)+y^2~g_2(\phi_1,\phi_2)=0.
\eqno(4.17)
$$
Then the ratios (4.13a,b) can be re-written as follows
$$
\exp((x_k-x_l)\phi_1+(x_k^2-x_l^2)\phi_2)=\frac{(u-x_k)}{(u-x_l)}\frac{(v-x_k)}{(v-x_l)}.\eqno(4.18)
$$
This equation is true for any $k,l=1,2,3,~k\neq l.$ This is to say, for any index we have same $\phi_1,\phi_2$ and same
$u,v$. Here $u,v$ depend of two parameters $\phi_1,\phi_2$.

Inversely, If we have $u=u(\varphi_u),~v=v(\varphi_v)$, then we
can find corresponding $g$ by
$$
\frac{g_0}{g_2}=uv,~~\frac{g_1}{g_2}=u+v.
$$
From this two equations we find $\phi_1$ and $\phi_2$. We expect
that
$$
\exp(V(\varphi_u+\varphi_v))=\exp(V\phi_{2}),
$$
or,
$$
\varphi_u+\varphi_v=\phi_{2}.
$$
In this way we have established connection between characteristics
of general complex algebra and solutions of Riccati-Abel equation.

The next task is to prove that the ratio $u=-g_0/g_1|_{g_2=0}$, in
fact, satisfies the Riccati-Abel equation.

Now, let us calculate derivatives of $g_1,g_0$ under the following
condition
$$
g_2(\phi_1,\phi_2)=0. \eqno(4.19)
$$
From this equation it follows that $\phi_1$ is a function of
$\phi_2$,  $\phi_1=\phi_1(\phi_2)$. Thus, we have to prove that
the function
$$
u(\phi_2)=-\frac{g_0(\phi_1(\phi_2),\phi_2)}{g_1(\phi_1(\phi_2),\phi_2)},
\eqno(4.20)
$$
obeys the Riccati-Abel equation.

Differentiating  equation (4.19) we obtain
$$
\frac{d g_2}{d \phi_2}+\frac{d g_2}{d
\phi_1}\frac{d\phi_1}{d\phi_2}=0. \eqno(4.21)
$$
From this equation taking into account constraint (4.21) we get
$$
\frac{d \phi_1}{d \phi_2}=-\frac{1}{g_1}(g_0+a_2g_1). \eqno(4.22)
$$
Now we have to use the following formulae
$$
\frac{d g_0}{d \phi_2}=a_0g_1+\frac{dg_0}{d\phi_1}\frac{d
\phi_1}{d\phi_2}
=a_0g_1-a_0\frac{g_2}{g_1}(g_0+a_2g_1)|_{g_2=0}=a_0g_1.
$$
$$
\frac{d g_1}{d \phi_2}=-a_1g_1-\frac{d g_1}{d \phi_1}\frac{d
\phi_1}{d \phi_2} =-a_1g_1-\frac{g_0}{g_1}(g_0+a_2g_1).
$$
By using these  formulae we are able to calculate derivative of
the fraction:
$$
\frac{d}{d \phi_2}\frac{g_0}{g_1}=\frac{g_0'g_1-g_0g_1'}{g_1^2}=
\frac{1}{g_1^3}(~a_0g_1^3+a_1g_1^2g_0+g_0^3+a_2g_1g_0^2).
\eqno(4.23)
$$
Coming back to definition (4.20) transform (4.23) into
Riccati-Abel equation:
$$
\frac{du }{d\phi_2}=-a_0+a_1u^2+u^3-a_2u^2. \eqno(4.24)
$$

{\bf Concluding remarks}.

As the ordinary Riccati equation, also the Riccati-Abel equation
has a relationship with linear differential equation. Seeking a
summation formula for solutions of Riccati-Abel equation we
established a certain interrelation between these solutions with
multi-trigonometric functions of third order. We have elaborated
some rule according to which in order to build a summation formula
for solutions of Riccati-Abel equations it is necessary to
consider the pair of solutions, which can be achieved by using an
auxiliary variable. This idea can be successfully used for the
solutions of generalized Riccati equations of any order with
constant coefficients. By increasing the order of the
non-linearity the number of auxiliary variables also will
increase. For example, from solutions of generalized Riccati
equations of fourth order we have to compose the triplet of
solutions with two auxiliary variables, and for $n$-order
generalized Riccati equations it is necessary to compose
$(n-1)$-pulet of solutions with $(n-1)$ auxiliary variables.


\begin{thebibliography}{99}



\bibitem{Riccati} Davis H.T. Introduction to nonlinear
differential and integral equations. United States Atomic Energy
Commission. U.S.Goverment Printing Office, washington
D.C.Reprinted Dover, New York 1960.
\bibitem{Abel} N.H.Abel,  Oeuvres completes du Niels Henrik Abel.-Christiana, 1881.
\bibitem{Yeomans}
Yeomans J.M. Statistical mechanics of phase transitions. Clarendon Press, Oxford. (1992).\\
John Cardy. Scaling and renormalization in statistical physics.
Cambridge lecture notes in physics. Eds. P.Goddard, J.Yeomans.
Cambridge university press 1996. ISBN 052149959 3.
\bibitem{Ungar1} A.Ungar, Addition theorems inordinary differential
equations. {\it Amer.Math.Monthly} {\bf 94} (1987) 872-875.
\bibitem{Ungar2}
 A.Ungar, Addition theorems for solutions to linear
homogeneous constant coefficient differential equations. {\it
Aequatios Math}. {\bf 26} (1983), 104-112.
\bibitem{Akhiezer}
  N. I. Akhiezer, Elements of the Theory of Elliptic Functions,
(1970) Moscow, translated into English as AMS Translations of
Mathematical Monographs Volume 79 (1990) AMS, Rhode Island ISBN
0-8218-4532-2
\bibitem{Yamaleev1} R.M.Yamaleev, {\bf Geometrical and physical interpretation of evolution governed by general complex algebra }
{\it Journal of Mathematical Analysis and Applications}
doi:10.1016/j.jmaa. (2007)09.018; 340(2008) 1046-1057
\bibitem{Babusci} Babusci D.,Dattoli G.,Di Palma E.,Sabia E.,
Complex-type numbers and generalization of the Euler identity.
{\it Adv.Appl.Clifford Al.} {\bf 22} (2012) 271.
\bibitem{fraction}
 Vedic Mathematics: Sixteen Simple Mathematical Formulae from the
Vedas, by Swami Sankaracarya (1884-1960), Motilal Banarsidass
Indological Publishers and Booksellers, Varnasi, India, 1965; \\
Stapel, Elizabeth. "Partial-Fraction Decomposition: General
Techniques."  Purplemath. Available from
http://www.purplemath.com/modules/partfrac.htm.
\bibitem{Vein} Vein R., Dale P., "Determinants and their
applications in mathematical physics", Springer-Verlag, New York,
Inc., 1999. ISBN 0-387-98558-1.
\bibitem{Yamaleev2} R.M.Yamaleev {\bf Multicomplex algebras on polynomials and generalized Hamilton dynamics } {\it Journal of Mathematical Analysis
and Applications} {\bf 322} (2006) 815-824.
\bibitem{Yamaleev3} R.M.Yamaleev {\bf Complex algebras on n-order polynomials and generalizations of trigonometry, oscillator model
and Hamilton dynamics } {\it J. Adv. Appl. Clifford Al.} {\bf 15}
No.2 (2005) 123-150.




\end{thebibliography}
\end{document}